\documentstyle[12pt,aaspp4]{article}
\input epsf

\begin{document}

\title{Dynamical Quenching of the $\alpha^2$ Dynamo}

\centerline{George B. Field$^1$ and Eric G. Blackman$^2$}
\bigskip

\centerline{1. Harvard-Smithsonian Center for Astrophysics, 60 Garden St., 
Cambridge MA, 02138}

\centerline{2. Department of Physics \& Astronomy, and Laboratory}
\centerline{for Laser Energetics, University of Rochester, Rochester NY, 14627}

\centerline{(submitted to ApJ)}

\def\be#1{\begin{equation} \label{#1}}
\def\pref#1{(\ref{#1})}
\def\ee{\end{equation}}
\def\beq{\begin{eqnarray}}
\def\eeq{\end{eqnarray}}
\def\nn{\nonumber \\ }
\def\ano#1{\eqno{({\rm A.#1})}}
\def\bno#1{\eqno{({\rm B.#1})}}
\def\bye{\end{document}}
\def\lb{\langle}
\def\rb{\rangle}
\def\ts{\times}
\def\curl{\nabla\ts}
\def\kms{\, {\rm km\, s}}
\def\pc{\, {\rm pc}}
\def\cms{\, {\rm cm\, s}}
\def\ed{{\rm ed}}
\def\G{\, {\rm G}}
\def\y{\, {\rm y}}
\def\const{{\it const}}
\def\exp{\, {\rm exp}\, }

\def\h{{\bf h}}
\def\k{{\bf k}}
\def\s{{\bf s}}
\def\bfj{{\bf j}}
\def\bS{{\bf S}}
\def\R{{\bf R}}
\def\G{{\bf G}}
\def\tfrac#1#2{{\textstyle{#1\over #2}}}
\def\half{\tfrac{1}{2}}

\def\O{{\cal O}}

\def\OE{\overline{\bf E}}
\def\obeta{\overline{\beta}}
\def\OA{\overline{\bf A}}
\def\oa{\overline{A}}
\def\ovar{\overline{\varphi}}
\def\J{{\bf J}}
\def\OJ{\overline{\J}}
\def\z{{\bf z}}
\def\x{{\bf x}}
\def\e{{\bf e}}
\def\A{{\bf A}}
\def\a{{\bf a}}
\def\H{{\bf H}}

\def\E{{\bf E}}
\def\ob{\overline{B}}
\def\OB{\overline{\bf B}}
\def\OV{\overline{\bf V}}
\def\lra#1{\left\langle #1\right\rangle}
\def\lrc#1{\left\{ #1\right\}}
\def\lrb#1{\left[ #1\right]}
\def\b{{\bf b}}
\def\bfb{{\bf b}}
\def\v{{\bf v}}
\def\bfv{{\bf v}}
\def\part{\partial}
\def\nt{\nabla\times}
\def\cnt{\cdot\nt}
\def\nab{\nabla}
\def\V{{\bf V}}
\def\la{\lambda}
\def\w{\omega}
\def\bet{\beta}
\def\lrp#1{\left(#1\right)}
\def\bff{{\bf f}}
\def\B{{\bf B}}
\def\etal{{\it et al.}\ }

 \baselineskip=24pt

\centerline{\bf Abstract}

We present a two-scale approximation for the dynamics of a nonlinear
$\alpha^2$ dynamo. Solutions of the resulting nonlinear equations
agree with the numerical simulations of Brandenburg (2001),
and show that $\alpha$ is quenched by the buildup of magnetic
helicity at the forcing scale $1/k_2$ as the $\alpha$ effect transfers it
from the large scale $1/k_1$. For times $t > (k_1/k_2)R_{M,2}$
in eddy turnover units  (where $R_{M,2}$ is the magnetic Reynolds
number of the forcing scale), $\alpha$  is resistively limited in the form
predicted for the steady-state case. However, for $t << R_{M,2}$,
$\alpha$ takes  on its kinematic value, independent of $R_{M,2}$,
allowing the production of large-scale magnetic energy equal to $k_1/k_2$
times equipartition. Thus the dynamic theory of $\alpha$
predicts substantial "fast" growth of large-scale field despite being
"slow" at large times.

\bigskip
{{\bf Subject Headings}:  MHD; turbulence; 
ISM: magnetic fields;  galaxies: magnetic fields; 
stars: magnetic fields; methods: numerical}

\section{Introduction}
Large-scale magnetic fields are often interpreted in terms of the equations
of Mean-Field Magnetohydrodynamics (Krause \& R\"adler 1980):
\be{n1}
\part_t\OB=\alpha\nt \OB+\lrp{\beta+\la} \nab^2 \OB\; ,
\ee
where $\OB$ is the mean (or large-scale) magnetic field,
\be{n2}
\la = {\eta c^2\over 4\pi}
\ee
is the magnetic diffusivity in terms of the resistivity $\eta$, and $\alpha$
and $\beta$ are parameters of the
underlying MHD turbulence.  Steenbeck \etal (1966) showed that if the
turbulence is isotropic and incompressible, and
the back reaction of $\OB$ is neglected,
\be{n3}
\alpha = \tfrac{1}{3} \tau\lra{\v\cdot\nt\v} \ee
and
\be{n4}\beta=\tfrac{1}{3} \tau \lra{v^2}\; .
\ee
Here $\tau$ is a typical correlation time of the flow $\v$, and
$\tfrac{1}{2}\lra{\v\cnt\v}$ is its kinetic
helicity, a measure of the net handedness of cyclonic motions (Parker 1955,1979).
$\beta$ represents turbulent diffusion of $\OB$.  
In this paper, the brackets and overbar represent spatial averages.

As $\OB$ grows, 
it exerts a backreaction on the turbulent flow,
and equations \pref{n3} and \pref{n4} must be modified to account for this.
A number of attempts to describe the corresponding saturation
of $\alpha$, or ``$\alpha$ quenching", have been
made.  As part of a general study of homogeneous, isotropic, helical
MHD turbulence, Pouquet, Frisch and L\'eorat (1976; hereafter PFL) used
the Eddy Damped Quasi-Normal Markov (EDQNM) approximation to derive
evolution equations for the spectra of kinetic
energy, magnetic energy, kinetic helicity, and magnetic helicity (defined as
$\lra{\A\cnt \A}/2$, with $\A$ the
vector potential).  They then solved a number of initial-value problems for
these spectra, and found an
$\alpha$ effect like that predicted by Steenbeck \etal (1966). 
By expanding in
terms of
a small quantity  $a$, they found that
$\alpha$ appropriate for a field having a large scale $k^{-1}$ is
\be{n5}
\alpha(k) = -\tfrac{2}{3}\int^\infty_{k/a} \theta_{kqq}
\lrp{H^V_q-H^C_q}dq\;, \ee
where $H^V_q$ is the spectrum of the small-scale kinetic helicity
\be{n6}
H^V=\half\lra{\v\cnt\v}\;,\ee
$H^C_q$  is the spectrum of the small-scale current helicity
\be{n7}
H^C = \half \lra{\b\cnt\b}\;,\ee
(where the small-scale field $\b$, like other magnetic fields in this paper,
 is in velocity units), and $\theta_{kqq}$
is the relaxation time for the interaction
of two wave numbers $q$ and $q'\sim q$ to excite $k\ll q$.  Equation
\pref{n5} is appropriate for the case that the
lower limit of $q$, $k/a$, is much larger than $k$, the wave number of the
large-scale field.
If one replaces $\theta_{kqq}$ by  $\tau$, the first term in
\pref{n5} agrees with \pref{n3}.  However,
the second term in \pref{n5} is new, and its physical significance was
discussed in PFL. It will play an
important role in what follows.

Gruzinov and Diamond (1994, hereafter GD, 1995, 1996) 
and Bhattacharjee and Yuan (1995, hereafter BY) 
recognized that the current 
helicity term in \pref{n5} and \pref{n7} is
related to magnetic helicity, a conserved quantity in ideal MHD, and
exploited that 
fact to find how $\alpha$ is quenched for a closed system 
when it has reached  a steady state.
In this paper, we also link  the current helicity contribution
to $\alpha$ with the equation for magnetic helicity evolution,
but in addition to considering  a steady state, we solve the 
time-dependent problem. As will be discussed,
the results ultimately lead to different conclusions
than those of GD and BY.

There is an important assumption built
into our approach: we assume that the PFL 
current helicity contribution to $\alpha$  
represents the current helicity to all orders 
in the mean field, not the zeroth-order quantities that appear
in the formalism of Field, Blackman and Chou (1999, hereafter FBC).
who expanded turbulent quantities about an isotropic state with $\ob=0$.
However, we have been unable to prove that this assumption is correct.
This issue is addressed in Blackman \& Field (1999), 
where it is shown that the current helicity that formally appears
as a correction in GD and BY should really be the 
zeroth-order contribution as in FBC.  It turns out, however, that the success
of the dynamical theory described below depends crucially
on ignoring this ordering ambiguity.

In section 2 we discuss the model of
PFL and produce a two-scale simplification of their equations.
(This is supplemented by Appendix A).
In section 3 we solve the resulting time-dependent equations 
for large-scale field growth and show that the results
agree well the numerical simulations of Brandenburg (2001, Hereafter B01).
In section 4 we compare our results to the implications
of previous $\alpha$ quenching models (and supplement this by Appendix B).
We conclude in section 5.

\setcounter{section}{1}

\section{Using PFL in a Two-Scale Approximation}
 
PFL studied the spectra of kinetic energy,
\be{n10}
E^V = \half \lra{v^2}\; , \ee
magnetic energy
\be{n11}
E^M = \half \lra{B^2}\; , \ee
kinetic helicity $H^V$ (eq.\ \ref{n6}), and magnetic helicity
\be{n12}
H^M = \half \lra{\A\cnt\A} = \half\lra{\A\cdot\B}\; . \ee
(Note the factor of $1/2$ in \pref{n12}.)
It is easy to show that the spectrum of current helicity (eq.\ \ref{n7}) is
related to that of $H^M$ by
\be{n13}
H^C_k = k^2H^M_k\; . \ee
Therefore the evolution of $H^C_k$, needed for evaluating $\alpha$
according to \pref{n5}, is tied to that of
$H^M_k$.

Magnetic helicity conservation will play a role in what follows.  Following
Moffatt (1978), we can use the induction
equation $(c=1)$ in the form
\be{n14}
\part_t \B = -\nt\E = \nt \lrp{\v\times \B-\la\nt \B} \ee
 to show that
\be{n15}
\part_t \lrp{\half \A\cdot\B} = -\half \nabla\cdot \lrb{\B(\phi-\v\cdot
\A)+\v(\A\cdot \B)} - \la \B\cnt \B\; . \ee
With appropriate boundary conditions on $\part V$, the average of the
divergence over the volume $V$ vanishes, and so
\be{n16}
\part_t H^M = -2\la H^C\; , \ee
showing that if $\la=0$, $H^M$ is conserved.  As shown by Moffatt (1978),
this expresses the fact that the linkage
between magnetic lines of force cannot change if they are frozen in the
fluid.  As we shall see, even though the total
magnetic helicity is conserved if $\lambda=0$, $\alpha$ causes it to flow from small
scales to large
scales.  

Equations (3.2) and (3.4) of PFL are
\be{n17}
\lrp{\part_t +2\la k^2} E^M_k = k\Gamma_k \lrp{E^V_k-E^M_k} + 2\alpha(k)
H^C_k - 2\beta(k) k^2E^M_k
\ee
and
\be{n18}
\lrp{\part_t +2\la k^2} H^M_k = {\Gamma_k\over k} \lrp{ H^V_k-H^C_k} +
2\alpha(k)E^M_k-2\beta(k)H^C_k\; . \ee
We have restored the Ohmic dissipation term $2\la k^2$ which appears in PFL
Table 1, but which is omitted in PFL
(3.2) and (3.4).  We also have omitted $\hat{\Gamma}_k$, which PFL state is
a first-order correction to $\Gamma_k$.

The
numerical results of PFL show that if helical MHD turbulence is excited predominately at a single wave number $k_2$ (the outer scale of the
turbulence), a pulse of excitation at $k_1$, say, moves toward smaller
values (larger scales) in what is
sometimes called an inverse cascade.  B01 has shown
that this inverse cascade is non-local in the sense that the excitation
jumps from $k_2$ to $k_1$.

We are interested in applying \pref{n18} to the case $k=k_1$.  Because according to (5)
$\alpha (k_1)$
is based on helicity at $k\gg k_1$, and
$E^M(k_1)$ can be significant, the $\alpha$-effect pumps magnetic
helicity from $k_2$ to $k_1$.  As we will see later, if
$\la$ is small magnetic helicity conservation requires that an equal and
opposite amount of helicity must be established at
$k_2$.  Because of \pref{n13}, the last term of \pref{n18}, 
$(2\beta(k) H^C_k)$ has the
same qualitative effect as $2\la k^2H^M_k$.
The first term on the right is
proportional to
\be{n19}
\Gamma_k = \tfrac{4}{3}k \int^{ak}_0 \theta_{kkq}E^M_qdq\; , \ee
which depends on the magnetic energy at $k\ll k_1$. In the
numerical results of PFL  there is a
peak at $k_1$, with little energy at $k<k_1$.  Hence we assume that the 
$\Gamma_k$ term is negligible.
Note that $\alpha(k)=\half \alpha^R$ and $\beta(k)=\nu^V_k$ of PFL.

We may integrate 
of \pref{n17} and \pref{n18} over $k$, and approximate the results by
\be{n20}
\part_tE^M_1 = 2\alpha k^2_1H^M_1 - 2(\la+\beta) k^2_1 E^M_1 \ee
and
\be{n21}
\part_t H^M_1 = 2\alpha E^M_1-2(\la+\beta) k^2_1 H^M_1\; , \ee
where
\be{n22}
E^M_1 = \int_{q\sim k_1} E^M_q dq\ee
etc., in effect setting $\alpha(k_1)$ and $\beta(k_1)$ equal to their values 
derived from contrubutions at $k=k_2$.  (Note that $\alpha$ is dimensionally
a speed and $\beta$ a diffusivity. 
For magnetic and kinetic spectrum approximately Kolmogorov,
the dominant contribution to  both $\alpha$ and
$\beta$ comes from the forcing scale.)
This  two-scale approach, wherein the forcing scale is equal to
the scale at which the small scale field is peaked, is justified 
only when the forcing is sufficiently helical (Maron \& Blackman 2002).

It is reassuring that \pref{n20} and \pref{n21} are exactly the equations
one gets from two-scale theory applied to
the small scale $k_2$ and the large scale $k_1$ (Appendix A).  In what
follows we will often use $E^M_1=B^2_1/2$, where
$B_1$ is the field at scale $k^{-1}_1$.  Note that $B_1$ can be used
interchangeably with $\ob$.

In the same spirit, we can replace $\theta_{kqq}$ in \pref{n5} by a typical
value $\tau$ related to the peak at $k_2$, to
obtain
\be{n23}
\alpha = -\tfrac{2}{3}\tau \lrp{H^V_2-H^C_2}\; .\ee

\section{Dynamical quenching and comparison to numerical simulations}

Here we show that the solutions of the equations in the  
two-scale formalism of the previous section 
agree well with the numerical results of \cite{b2001}.
B01 studied the dynamo effect in a nearly
incompressible conducting fluid with periodic
boundary conditions.   $B_1$ is   allowed to grow at various wave numbers
$k$, consistent with the boundary conditions,
as a result of the $\alpha$ effect, thus  simulating  a non-linear 
$\alpha^2$ dynamo.
The results of B01 are qualitatively similar to the 
numerical results of PFL, in
that a pulse of excitation propagates to large
scales.  As B01 kept $H^V_2$ approximately constant by driving the MHD
turbulence with a helical force, in the light of \pref{n23},
  \pref{n20} and \pref{n21} become
\be{n37}
\part_t E^M_1 = 2 \lrp{\alpha_0 +\tfrac{2}{3}\tau k^2_2H^M_2} k^2_1H^M_1 -
2(\la+\beta) k^2_1 E^M_1 \ee
and
\be{n38}
\part_t H^M_1 = 2 \lrp{\alpha_0 +{\tfrac{2}{3}}\tau k^2_2
H^M_2}E^M_1-2(\la+\beta)k^2_1H^M_1\;,\ee
where
\be{38p}
\alpha_0 = -\tfrac23\tau H^V_2 = {\it const.} \ee
Remarkably, the nonlinear differential equations \pref{n37} and \pref{n38}
have a force-free solution for $B_1$ in which
\be{n39}
E^M_1=k_1H^M_1\; . \ee
Thus, \pref{n38} and \pref{38p} reduce to one
equation that $H_1$ must satisfy, namely 
\be{n40}
\part_t H^M_1 = 2k_1 \lrp{\alpha_0 + \tfrac{2}{3}\tau k^2_2
H^M_2}H^M_1-2(\la+\beta)k^2_1 H^M_1 \; .\ee
To solve \pref{n40} and \pref{40p}, we 
express $H^M_2$ in terms of $H^M_1$.  To
do this, we use the conservation of magnetic helicity, \pref{n16}, 
in the form
\be{40p}
\part_t H^M_1+\part_t H^M_2 = -2\la \lrp{k^2_1 H^M_1 +k^2_2 H^M_2}\; .
\ee
Equations 
\pref{n40} and \pref{40p} are the coupled equations 
in $H^M_1$ and $H_2^M$ which need to be
solved. 
If $H_2^M$ is small, $H_1^M$, and hence the large-scale field $B_1$
grows exponetially, driven by the first term on the right hand
side of \pref{n40}. The magnetic helicity conservation equation \pref{40p} 
shows that, for small $\lambda$, 
growth of $H_1^M$ is not free, but comes at the
expense of growing $H_2^M$ with the opposite sign.  
This decreases the value of $\alpha$ in \pref{n40}.
This ``$\alpha$-quenching'' slows the growth of 
$H_1^M$, leading to a steady state when the right-hand side
of $\pref{n40}$ vanishes.

To solve \pref{n40} and \pref{40p}, we rewrite them in dimensionless form.
We define the dimensionless magnetic helicities
 $h_1\equiv  2 H^M_1k_2/v_2^2$ and $h_2 \equiv 2 H^M_2 k_2/v_2^2$ and 
write time in units of $1/k_2v_2$. We also define  
$R_M\equiv (v_2/k_1)/\lambda$.
(Note that this definition of $R_M$ is based on the forcing-scale
RMS velocity but on the large scale, $k_1^{-1}$.  We will later employ
a second magnetic Reynolds number $R_{M,2}\equiv R_M (k_1/k_2)$.)
We also need a prescription for $\alpha_0$ and for $\beta$.
We assume that the kinetic helicity is forced maximally, and
we take $\tau=2/k_2v_2$ implying that $\alpha_0= 2v_2/3$. 
Unfortunately, a rigorous prescription for $\beta$ in 3-D is lacking, but 
as in B01, we will consider two cases, 
$\beta=\beta_0\alpha/\alpha_0$
and $\beta=\beta_0\equiv v_2/k_2$.

Using the above scalings we can replace 
(\ref{n40}) and (\ref{40p})
with dimensionless equations given by
\beq
\partial_t h_1 ={4\over 3}\left({k_1\over k_2}\right)h_1(1+h_2)-2h_1
\left[{k_1\over k_2 R_M}+{k_1^2\over k_2^2}(1+q_2h_2)\right]
\label{40b}
\eeq
and
\beq
\nn
\partial_t h_2 =-{4\over 3}\left({k_1\over k_2}\right)h_1(1+h_2)+2h_1
{k_1^2\over k_2^2}(1+q_2 h_2)
-{2\over R_M}{h_2k_2\over k_1},
\label{41b}
\eeq
where $q_2=0$ in the above equations corresponds to $\beta(t)=\beta_0=$constant.
and $q_2=1$ corresponds to $\beta(t)=\alpha(t)\beta_0/\alpha_0$.  
Solutions of these coupled equations are shown
in Figs. 1-4. The key parameters 
are $k_2/k_1$, $R_M$, and $q_2$. 
In the figures, we have also compared
these results to the empirical fits of numerical simulations 
in \cite{b2001}. We have taken $h_1(t=0)=10^{-3}$,
but the sensitivity to $h_1(0)$ is only  
logarithmic (see \pref{tkin} below).
In Fig. 1, we have used $k_2/k_1=5$, following B01, 
and in Fig. 2 we have used $k_2/k_1=20$.

In the figures, the solid lines represent our numerical 
solutions to \pref{40b} and \pref{41b}, whereas 
the dotted lines represent the formula given 
in \cite{b2001}, which is an 
empirical fit to simulation 
data assuming that $\alpha$ and $\beta$ are prescribed according
to \pref{48} and \pref{49} below.
More explicitly, B01 found that the growth of $\OB$ was
well described by the formula 
\beq
{B_1^2/B_{1,0}^2\over (1-B_1^2/B^2_{1,sat})^{1+{\alpha_0k_1-k_1^2\beta_0\over 
\lambda k_1^2}}}=
e^{2(\alpha_0k_1-k_1^2\beta_0)t}, 
\label{axel1}
\eeq
where $B_{1,0}= B_1(t=0)$.
This can be rewritten using the notation above as 
a dimensionless equation for $t$ in units of $(k_2v_2)^{-1}$, namely 
\beq
t={k_2\over 2k_1}{ {\rm Ln}[(h_1/h_0)(1 - h_1 k_1^2/k_2^2)^{R_M(k_1/k_2-2/3)-1}]
\over {2 / 3} - {k_1/k_2}}.
\label{axel2}
\eeq
Note that \pref{axel1} and \pref{axel2} 
correspond to $\alpha$ and $\beta$ quenching of the form 
\beq
\alpha={\alpha_0\over  1+s_B B_1^2/v_2^2}
\label{48}
\eeq
and 
\beq
\beta={\beta_0\over  1+s_B B_1^2/v_2^2},
\label{49}
\eeq
where 
$s_B\sim R_M (k_1/k_2)(2/3-k_1/k_2)=R_{M,2} (2/3-k_1/k_2)$,
and $R_{M,2}\equiv v_2/k_2\lambda$.
Eqns. \pref{48} and \pref{49} are derived from those in B01 by
re-scaling Eq. (55) of B01 with our notation.  
It can also be shown directly that, 
up to terms of order $1/R_M$,  
 \pref{axel2} is consistent 
with that derived by substituting \pref{48} and \pref{49} 
into \pref{40b} and solving for $t$.
Note that in contrast to the suggestion of B01, it is 
actually the forcing-scale magnetic Reynolds number, $R_{M,2}$, that plays 
a prominent role in these 
formulae.

The solutions of \pref{40b} and \pref{41b} 
are subtle and interesting. Some insight
can be gained by their sum
\beq
\nn
\partial_t h_1 + \partial_t h_2 
= -{2\over R_M}\left({h_1k_1\over k_2} +{h_2k_2\over k_1}\right),
\label{41bb}
\eeq
which corresponds to 
\pref{40p}, the conservation of total magnetic helicity. 
If we make the 
astrophysically relevant assumption that $R_M >> 1$, the right hand side of
\pref{40p} is small for all $h_1$ and $h_2$. It follows that 
$\partial_t (h_1 + h_2)=0$ 
and for $h(t=0)=0$,  this implies $h_2 = - h_1$. 
In this period, we can self-consistently ignore $1/R_M$ in
\pref{40b}. If $q_2 = 1$, this phase ends 
when $h_2= -1$, so that $h_1 = 1$. 
This is manifested in figure 3.
    
This kinematic phase precedes the asymptotic saturation of the dynamo 
investigated by other authors, in which all time derivatives
vanish exactly. For this to happen, the right hand side of \pref{41bb} must
vanish, which is equivalent to demanding that $h_2 = -(k_1/k_2)^2 h_1$. 
Since the right hand sides of \pref{40b} and \pref{41b} 
are proportional to $1 + h_2$ when terms
of order $1/R_M$ are neglected, their vanishing requires that $h_2 = -1$, 
and therefore, that $h_1 = (k_2/k_1)^2$. This is observed in figures 1 and 2.
The asymptotic saturation  (when the field growth ceases)
takes a time of order $t_{sat} \sim  R_M k_2/k_1$, 
which in astrophysics is often huge. Thus, although in principle it is correct
that $\alpha$ is resistively limited 
(as seen from our solutions in figures 4 \& 5)
as suggested by BY, GD, Vainshtein \& Cattaneo (1992) and 
Cattaneo \& Hughes (1994), this is 
less important than the fact that for a time $t_{kin} < R_M$ 
the kinematic value of $\alpha$ applies.
The time scale $t_{kin}$ here is given by
a few kinematic growth time scales for the $\alpha^2$ dynamo,
more specifically, 
\beq
t_{kin}\sim {\rm Ln}[1/h_1(0)] (k_2/k_1)/(4/3-2k_1/k_2).
\label{tkin}
\eeq
For $h_1(0)=0.001$, $k_2/k_1=5$, $t_{kin}\sim 37$, as seen in Fig 3.
 
Note that $t_{kin}$ is sensitive to $k_2/k_1$
and independent of $R_M$.  Figure 3 shows that there is significant
disagreement in this regime with \pref{48},
but this formula was used in B01 only to model the  regime $t>R_M$, 
so the result is not unexpected. 
We can see from the solution  for $\alpha$ itself
that indeed our solutions do match \pref{48} 
for  $t > R_M$ (figures 4 and 5).  
Figure 4 shows the difference in the $\alpha$ along with \pref{49}
for the two values $R_M=10^2$ and $R_M=10^3$. Notice again
the disagreement with the  formula \pref{48}
until $t=R_M$, and agreement afterward.  
This marks the time  at which the resistive term
on the right of \pref{40b} becomes competitive with the  
terms involving $(1+h_2)$. Asymptotic saturation 
does not occur until $t\sim t_{sat}=R_M k_2/k_1$ as described above.


Finally, note that $q_2$ corresponds to 
$\beta=\beta_0$.  
In general, this leads to a lower value of $h_1$ in the asymptotic 
saturation phase because 
this enforces zero saturation of $\beta$, whereas there is
still some saturation of $\alpha$ in this limit.
(Note that $q_2=0$ corresponds to the case of GD discussed further
in appendix B.)
For large $k_2/k_1$ the solutions of \pref{40b}
and \pref{41b} are insensitive to $q_2=0$ or $q_2=1$. This is because
the larger $k_2/k_1$, the smaller the influence of the
$q_2$ terms in \pref{40b} and \pref{41b}.  This is highlighted 
in figure 6 where the result for $q_2=0$ is plotted with the B01
fit. This suggests that for large-scale separation, the magnetic energy
saturation is  insensitive to the form of $\beta$ quenching. 
However, in real dynamos, magnetic flux and not just magnetic energy
may be needed, so the insensitivity can be misleading because
$\beta$ is needed to remove flux of the opposite sign.
From the low $k_2/k_1$ cases, it is clear that that $q_2=1$
is a better fit to the simulations of \cite{b2001}.

\vspace{.1cm} \hbox to \hsize{ \hfill \epsfxsize8cm
\epsffile{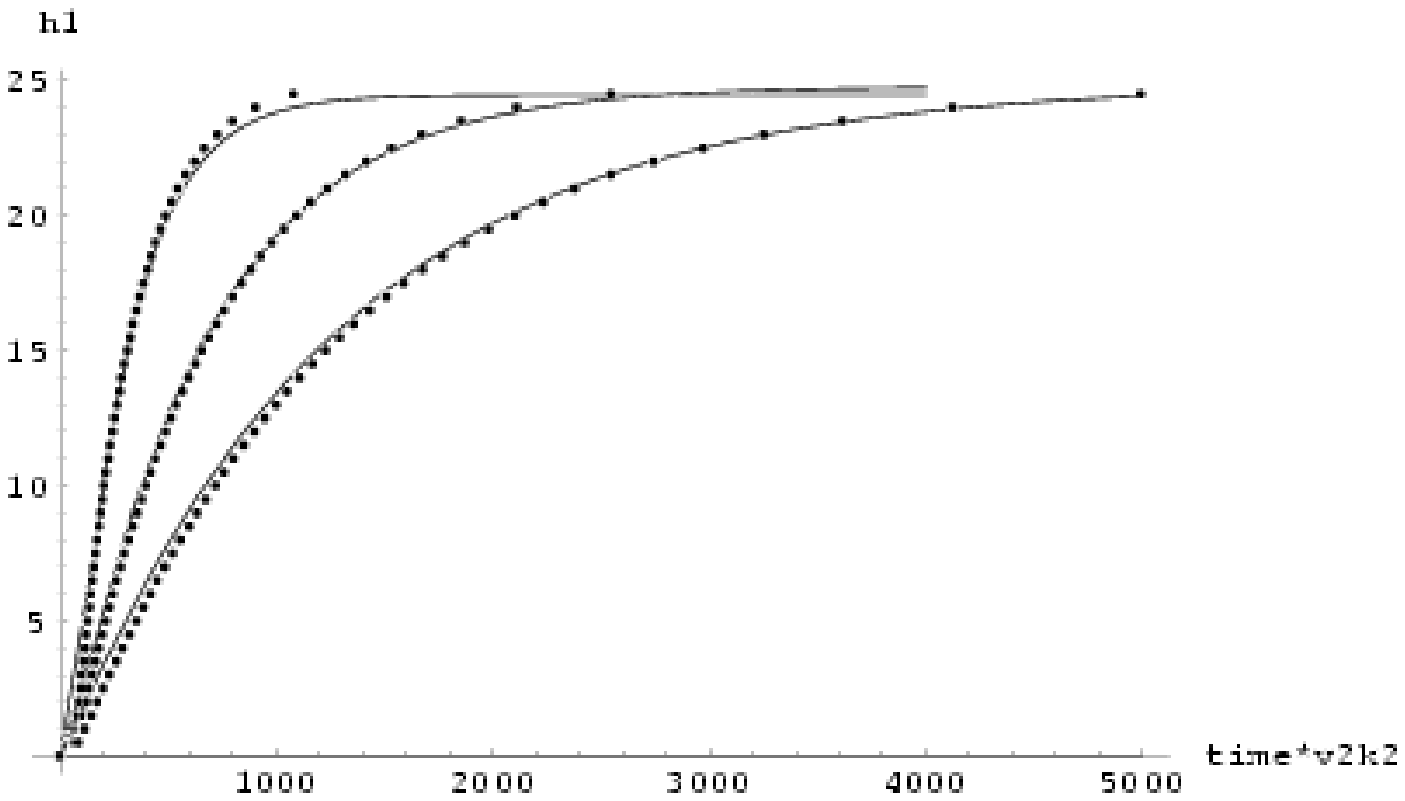} 
\hfill } \noindent 
{Figure 1: Solution for $h_1(t)$, $f_h=1$,$q_2=1$. Here $k_2/k_1 =5$ and
the three curves from left to right have $R_M=100,250,500$
respectively. The dotted lines are plotted from the formula used to 
quasi-empirically fit the simulations in B01 as described
in the text.

\vspace{.1cm} \hbox to \hsize{ \hfill \epsfxsize8cm
\epsffile{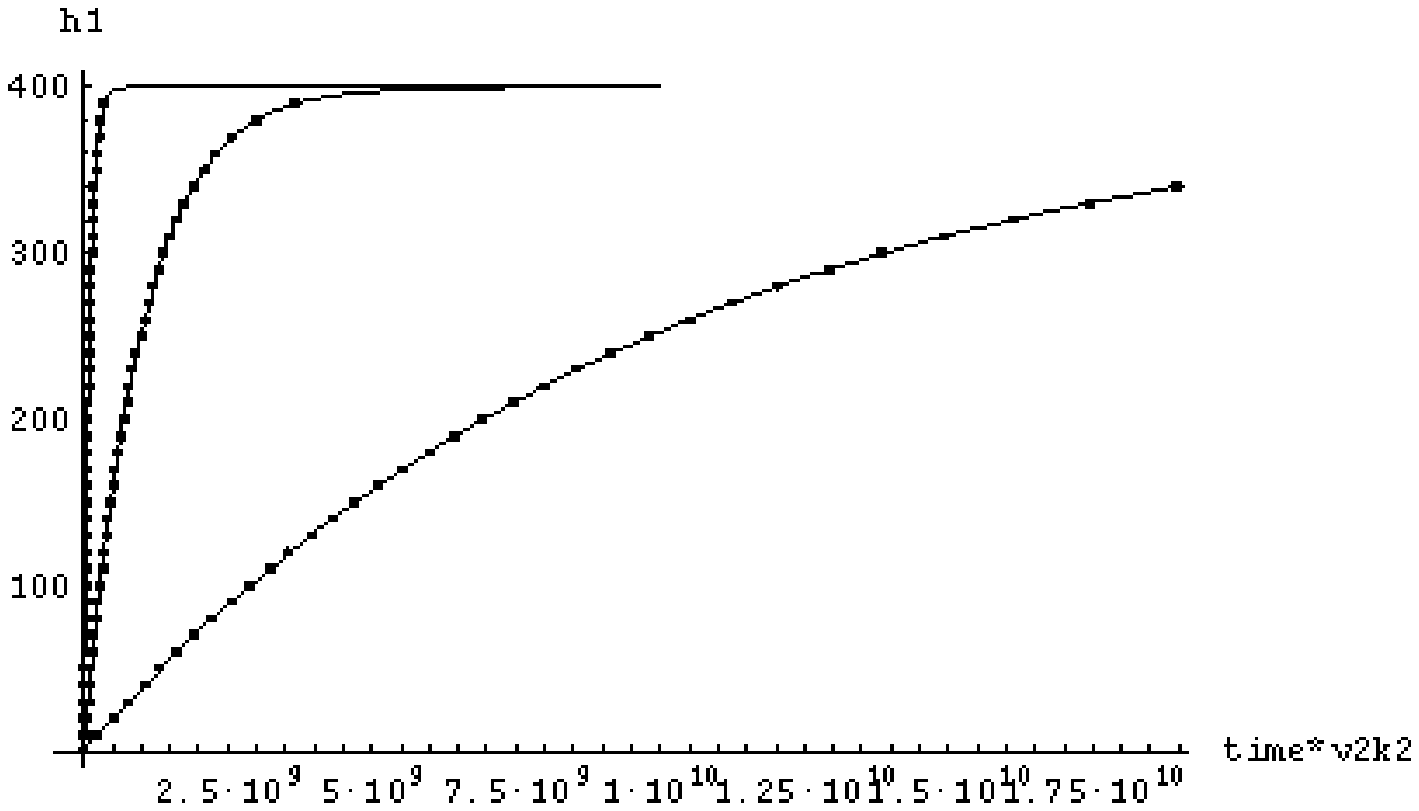} 
\hfill } \noindent 
{Figure 2: Solution for $h_1(t)$, $f_h=1$, $q_2=1$. Here $k_2/k_1 =20$ and
the three curves from left to right have $R_M=10^7,10^8,10^9$
respectively. The dotted lines are plotted from the formula for used to quasi-empirically fit simulations of B01.

\vspace{.1cm} \hbox to \hsize{ \hfill \epsfxsize8cm
\epsffile{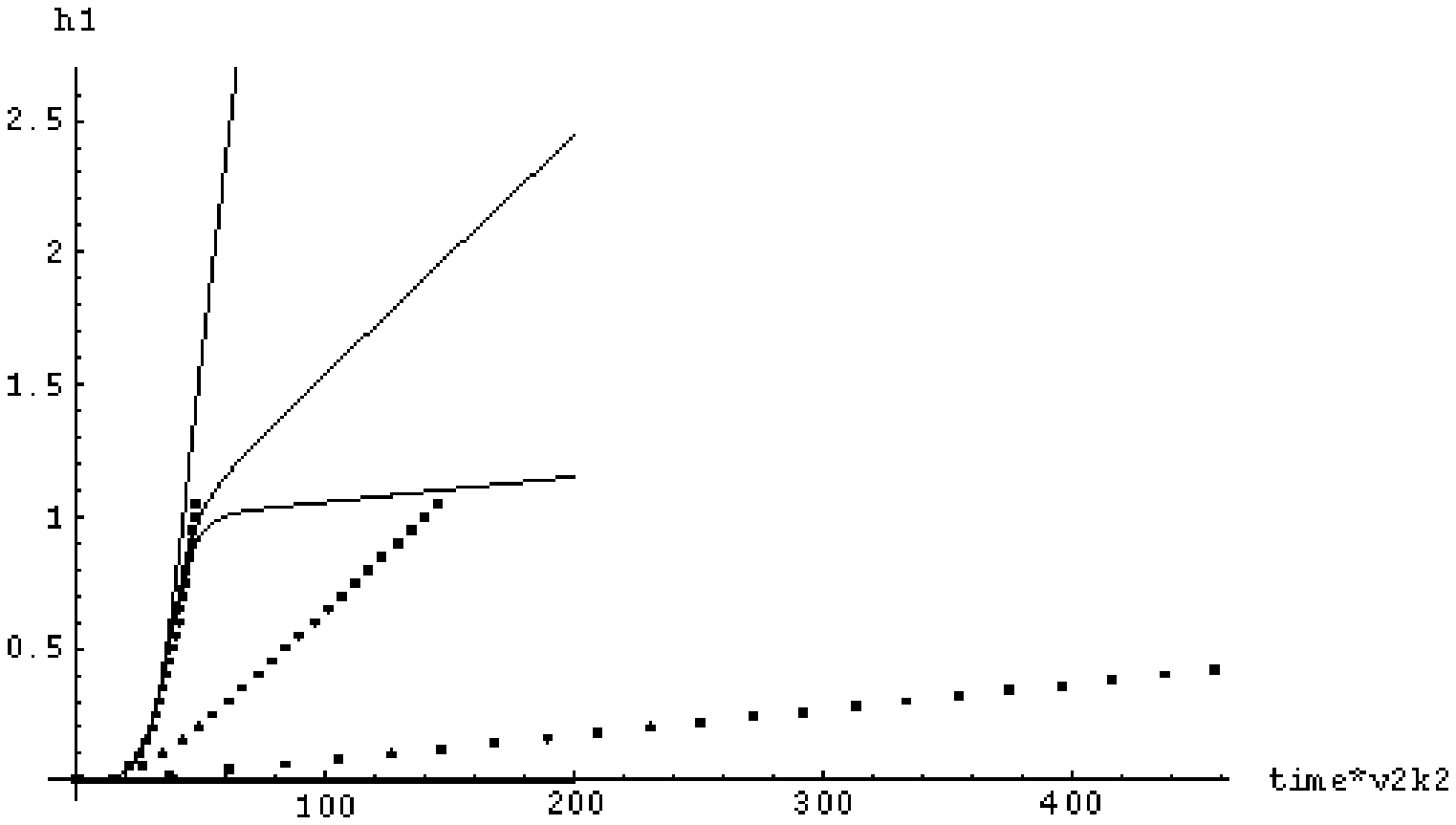} 
\hfill } \noindent 
{Figure 3: The early-time solution for $h_1(t)$, $f_h=1$, $q_2=1$. Here  
for $k_2/k_1=5$, and $R_M=10^2,10^3, 10^4$ from left to right respectively.
Notice the significant departure from the 
formula of B01 at these early times. For $t< t_{kin}$ there
is no dependence on $R_M$ and the growth proceeds kinematically.}

\vspace{.1cm} \hbox to \hsize{ \hfill \epsfxsize8cm
\epsffile{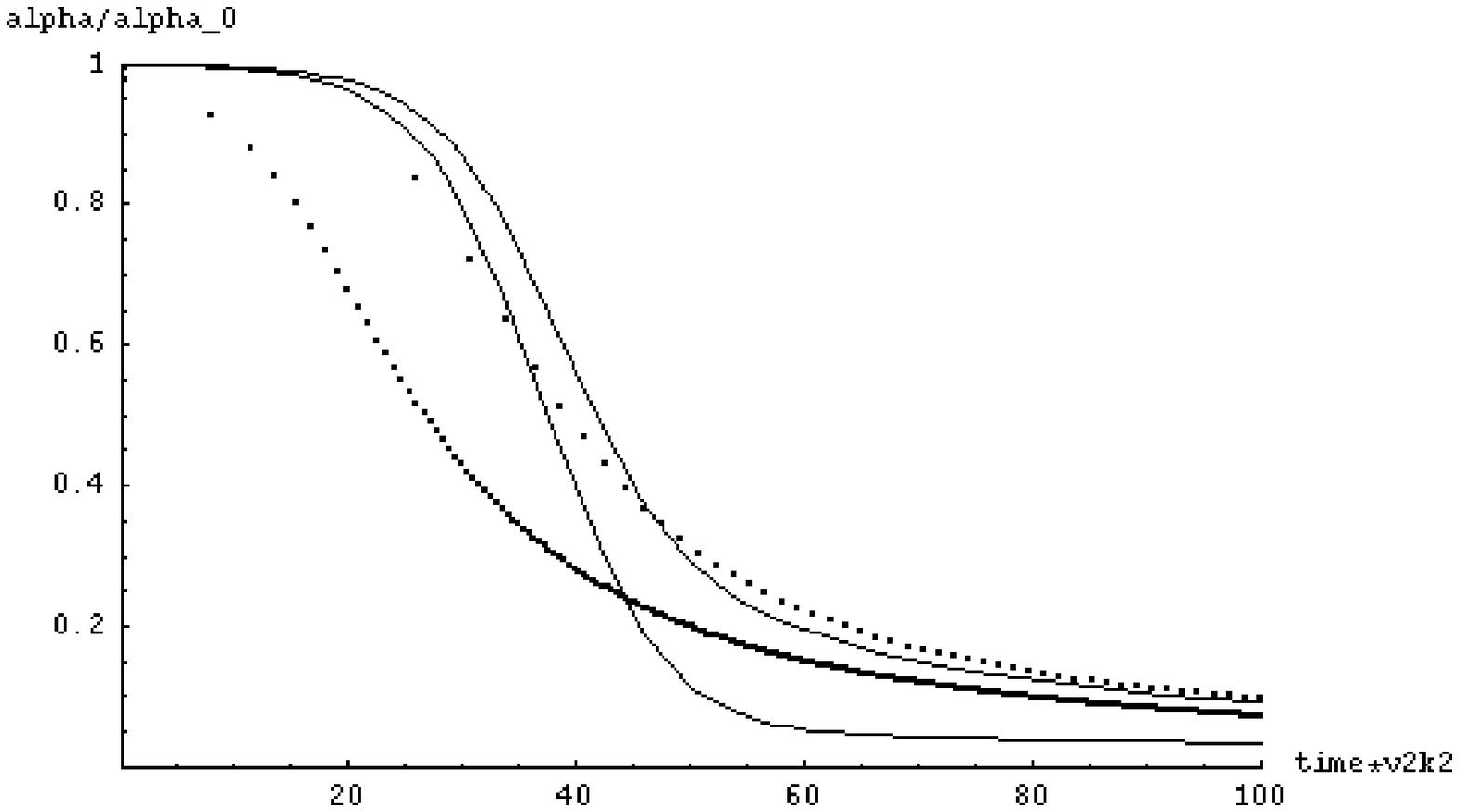} 
\hfill } \noindent 
{Figure 4: Solution of $\alpha/\alpha_0(t)$ 
for $h_1(t)$, $f_h=1$, $q_2=0$. Here $k_2/k_1 =5$ and
the solid lines are our solutions for $R_M=10^2$ (top curve)
and $R_M=10^3$ (bottom curve) respectively.
The top and bottom dotted curves are  from \pref{48}, interpreted
from Ref. \cite{b2001}}. Notice the longer kinematic
phase for our solutions, the overshoot, and the  convergence
of the solution for $R_M=10^2$ with that of \pref{48} at $t=R_M$.

\vspace{.1cm} \hbox to \hsize{ \hfill \epsfxsize8cm
\epsffile{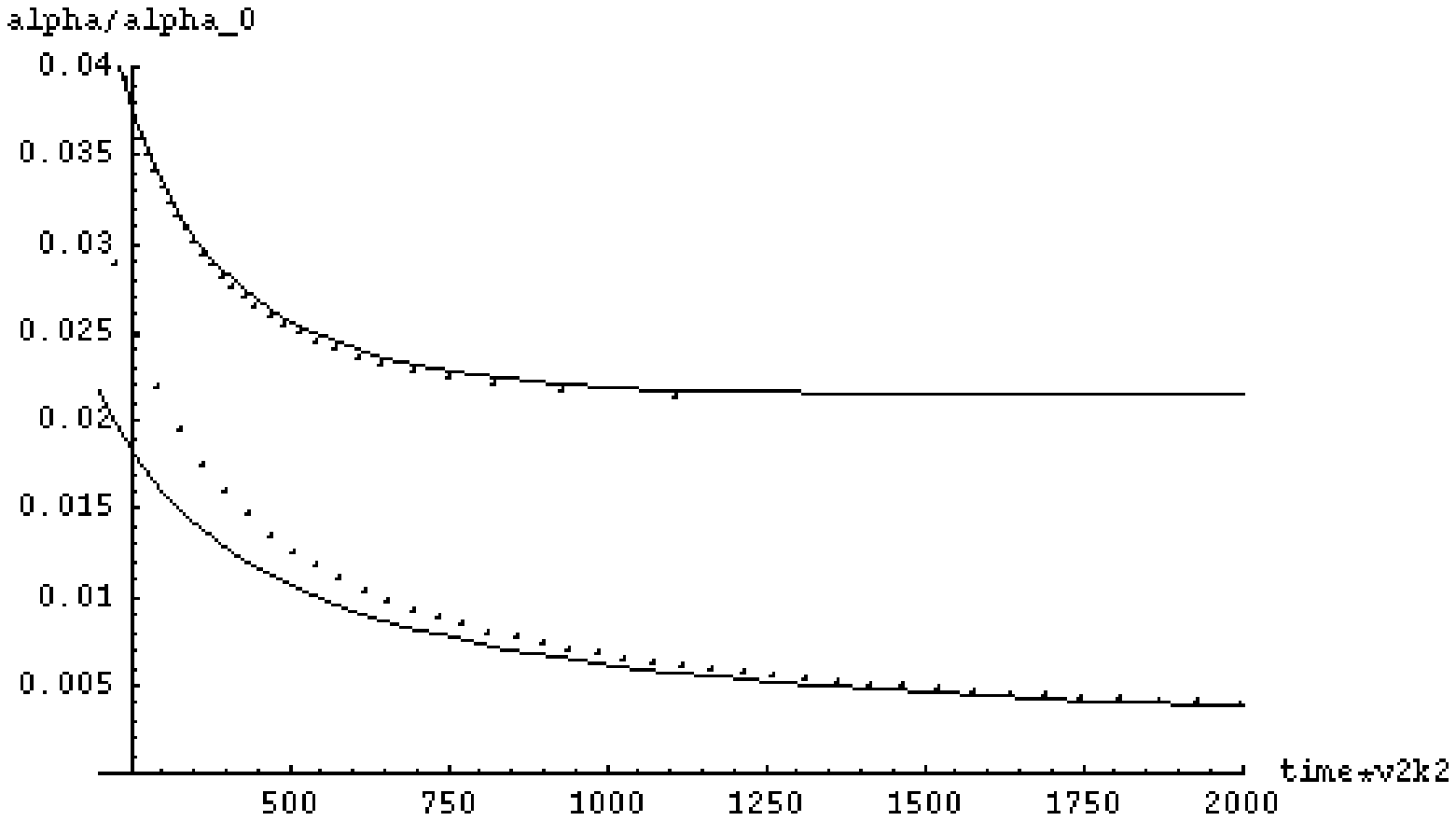} 
\hfill } \noindent 
{Figure 5: This is the extension of  Figure 5 for later times. Notice 
the convergence of the $R_M=10^3$ solution to \pref{48} near $t=R_M$.

\bigskip
\bigskip

Figure 6 (separate file) 
: Solution for $h_1(t)$, $f_h=1$, $q_2=0$. Here $k_2/k_1 =50$ and
the three curves from left to right have $R_M=10^2,10^3,10^4$
respectively. The dotted lines are plotted from the formula 
used to quasi-empirically fit simulations of Ref. \cite{b2001}}. For such 
large $k_2/k_1$ the fit to the data is only weakly sensitive to 
whether $q_2=1$ or $q_2=0$.

\bigskip
\bigskip

\section {Implications  and Comparison to Previous Work}

The physical picture of the quenching process just described
is this: helical turbulence is forced at $k_2$ ($=5$ in
B01), and kept approximately constant by forcing.  
Hence $\alpha_0 = -2\tau H^V_2/3 = {\it const}$.  
If $H_1^M$, the magnetic 
helicity at $k_1$ (which reaches $1$ here as a result of boundary
conditions), is initially small --- so that $|2k^2_1H^M_1/3|\ll |\alpha_0|$, \pref{n40} (or \pref{40b}) 
shows that it will be exponentially
amplified provided that the damping due to
$\beta +\la$ does not overcome the $\alpha$ effect.  
Initially, $\alpha=\alpha_0$, 
acting like a pump that  moves magnetic helicity from $k_2$ to $k_1$ 
and driving the dynamo.  This kinematic phase lasts
until $t_{kin}$ as given by \pref{tkin}. 
Eventually, the growing 
$H^M_1$
results in a  growing $H^M_2$ of opposite sign,
which reduces $\alpha$ through $H^C_2$.  
$R_M$-dependent quenching kicks in at $t=t_{kin}$, 
but it is not until $t=R_M$ that the asymptotic formulae 
\pref{48} and \pref{49} 
are appropriate.  Asymptotic saturation, defined by the time 
at which $B_1$ approaches 
its maximum possible value of $(k_2/k_1)^{1/2} v_2$,  
occurs at $t=t_{sat}=R_Mk_2/k_1$.
For $t\ge R_M$ our numerical solution, like the full numerical
simulations of B01,
is  well fit by the $\alpha$ in \pref{48} with a corresponding $\beta$
of \pref{49}. Our two-scale approach is also consistent with 
B01 in that magnetic helicity jumps from $k_2$ to $k_1$ without filling
in the intermediate wavenumbers.

The emergence of the time scale $t_{kin}$ is 
interesting because it shows how 
one can misinterpret the implications of the
asymptotic quenching formula \pref{48} and \pref{49}.
These formulae are appropriate only for $t> R_M$. 
The large-scale field actually grows kinematically
up to a value $B_1=(k_1/k_2)^{1/2} v_2$  by $t=t_{kin}$
and ultimately up to $B_1\sim (k_2/k_1)^{1/2}v_2$ by $t=t_{sat}$.
For large $R_{M,2}$, 
these values of $B_1$ are  both 
much larger than the quantity $v_2/R_{M,2}^{1/2}$, which 
would have been inferred to be the saturation value if 
one assumed 
\pref{48} and \pref{49} were valid at all times.


Dynamical quenching or time-dependent approaches recognizing the current
helicity as a contributor to $\alpha$ 
have been discussed elsewhere (\cite{zeldovich,kleeorinetal,kleeorinruz,kr}
see also Ji 1999; Ji and Prager 2001), 
but here we have specifically linked the  PFL $\alpha$ correction to the 
helicity conservation in a simple two-scale approach. 
Other quenching studies for closed systems such as
Cattaneo \& Hughes (1994)  and \cite{by} advocated
values of $\alpha$ which are resistively limited
and of a form in agreement with \pref{48} 
but with the assumption of a steady  $B_1$.
Assuming \pref{n23}, and using \pref{n40} and \pref{40p} in the steady
state, their formulae can be easily derived. 
However, one must also have a prescription for $\beta$.  
If $\beta$ is proportional to $\alpha$, then formulae like
\pref{48} and \pref{49} emerge.  If  $\beta(t)=\beta_0$, 
as in GD, then a formula for $\alpha$
$without$ resistively limited quenching emerges (this 
requires a re-interpretation of their formulae--see appendix B).  
On the other hand, we have shown that for large $k_2/k_1$, the 
dynamo quenching is largely insensitive to $\beta$.

An important point to re-emphasize  is
that even when resistive quenching formulae are found 
from steady-state analyses, this does not 
necessarily  reflect the saturation value of $B_1$.
The fact that there exists a kinematic regime up until $t_{kin}$ means that by
the time formulae like \pref{48} and \pref{49} 
are valid, the field may have already grown substantially, as we have shown.
That being said, all of our analysis here is for the growth of 
magnetic energy to saturation for the simple $\alpha^2$ dynamo  in a box
as in B01.
In such  a system, 
the magnetic energy can saturate at super-equipartition values 
because it is force free. The extent to which these 
idealized studies apply to 
real astrophysical systems with boundaries,
shear, and stratification, or to dynamo cycle periods, 
still remains to be seen.

\section{Conclusion}

We have shown that the evolution equations of PFL, together with their 
formula for $\alpha$, leads to dynamical $\alpha$
quenching from the $\alpha$-induced flow of magnetic helicity
from small to large scales; the associated buildup of small-scale 
current helicity of the opposite sign eventually 
suppresses $\alpha$. This simple $\alpha^2$ dynamo process 
can be modeled using a two-scale formalism. We have identified
a time scale $t_{kin}$ up to which the dynamo
in a periodic box operates independently of $R_M$ and grows
to large values, of order  $\sim (k_1/k_2)^{1/2} v_2$.
At later times, the dynamo becomes slow.  
The dynamo coefficients become 
resistively limited, depending strongly on $R_M$.  
Our solutions agree with the numerical simulations of B01
for the regime of $t>R_M$, where B01
showed that \pref{48} fits the data.

{\bf Acknowledgments}

\noindent We thank the ITP at UC Santa Barbara 
and its participants for stimulating interactions during the
Spring 2000 workshop on astrophysical turbulence. In particular, we thank 
A. Brandenburg, S. Cowley, R. Kulsrud, J. Maron, and P. Diamond  
for discussions.  We also warmly thank the participants
of the memorable Virgin Gorda MHD turbulence Workshop of Dec 2001, namely
A. Brandenburg, R. Kulsrud, J. Maron, B. Mattheaus, and A. Pouquet 
for the intense and extended discussions which have directly influenced
the revised version of this paper.
EB acknowledges support from DOE grant DE-FG02-00ER54600.


\def\cno#1{\eqno{({\rm C.#1})}}

\section*{Appendix A: The Equivalence of PFL
and Two-Scale Theory}
\appendix  Here we show that \pref{n20} and \pref{n21} also follow from
two-scale theory.
Multiplying (1)  by $\OB$ gives
$$
\begin{array}{rcl}\part_t \lrp{\half \ob^2} &=& \alpha\OB\cdot\nt \OB -
(\la+\beta)\OB\cdot\nab^2\OB \\ [6pt]
&=& \alpha \OB\cdot\nt \OB+(\la+\beta) \OB\cdot \nt \nt \OB \\ [6pt]
&\doteq&\alpha\OB\cdot\nt \OB - (\la+\beta) (\nt \OB)^2\; , \end{array}
\ano{1}
$$
where
$\doteq$ means equal to within a divergence; from \pref{n15}, it is the true
equality for the indicated boundary conditions
.  Now let $k^{-1}_1$ be the scale of $\ob$.  Then
\beq
\OB\cnt \OB &=& \OB_1\cnt\OB_1\, \nn
&=& 2H^C_1 = 2k^2_1H^M_1\; , \nonumber
\eeq
and 
$$\lrp{\nt\OB}^2  \;  = \; 2 k^2_1 E^M_1 \; . \hskip.6in\ano{2}
$$
Thus (A.1)  becomes
$$
\part_t E^M_1 = 2\alpha k^2_1H^M_1-2(\la+\bet)k^2_1E^M_1 \ano{3}
$$
in agreement with \pref{n20}.

From \pref{n15}
$$
\part_t \lrp{\half\OA \cdot\OB} \doteq - \OE\cdot\OB\; . \ano{4}
$$
Since from \pref{n1} and \pref{n14}
$$\hskip.6in \OE = -\alpha\OB+(\la+\beta)\nt \OB\; ,\ano{5}$$
$$ \part_t \lrp{\half\OA\cdot\OB}  \doteq  \alpha \ob^2 -
(\la+\beta)\OB\cnt\OB \ano{6}$$
  or 
$$
\hskip.4in \part_t H^M_1  \doteq  2\alpha E^M_1 - 2(\la+\beta) k^2_1 H^M_1\;
, \ano{7}
$$
in agreement with \pref{n21}.  We conclude that if the Alfv\'en effect is
omitted, and the field is concentrated at $k_1$ and $k_2$, equations (3.2)
and (3.4) of PFL are equivalent to the two-scale
approximation.


\def\dno#1{\eqno{({\rm B.#1})}}
\section*{Appendix B\\ Reinterpretation of GD Quenching Formula}


GD (1994) were the first to use the conservation of magnetic helicity to
obtain a formula for $\alpha$ in a closed system in a
steady state.  Their conclusion that $\alpha$ saturates when $B_1$ is of the
order of $R^{-1/2}_m$ $v_2$ stimulated the
present investigation, for if correct, it would imply that the $\alpha$
effect in the Galaxy would be useless in explaining any
fields larger than $10^{-16}$ Gauss, as $R_m\cong 10^{20}$ in the
interstellar medium.
Taking their formula for $\alpha$, 
we show that their result was misinterpreted.

GD did not assume that $\B_1$ is constant in space, so we can
use \pref{n21} with $k_1\ne 0$.  Because they
assumed a steady state, we put $\part_t=0$, so that
$$
\alpha E^M_1 = \la H^C_1 +\beta H^C_1\; , \dno{1}
$$
which agrees with eq.\ (9) in GD (1994) (except for a sign error in the
latter which is not propagated in the rest of their
paper.)

From \pref{n23} and \pref{38p} we have
$$
\alpha= \alpha_0 + \tfrac23 \tau H^C_2\; , \dno{2}
$$
so
$$
H^C_2 = \frac{3}{2\tau} (\alpha-\alpha_0)\; . \dno{3}
$$
If  $t>t_{sat}$
$$
H^C_1 = -H^C_2 = -{3\over 2\tau} (\alpha-\alpha_0) \; . \dno{4}
$$
If we substitute (B.4) into the first term on the right-hand side of (B.1),
we get
$$
\alpha = {\alpha_0 +\beta_0 \R\cnt \R\over 1+R^2}\; , \dno{5}
$$
where
$$
\R = \left({\tau\over 3\la}\right)^{1/2} \B_1 = {{\sqrt 2}R^{1/2}_{M,2}\over v_2}\B_1\; , \dno{6}
$$
and we have used $\tau = 2/k_2 v_2$ and $\la = \beta_0 /R_{M,2}=v_2 /3k_2 
{R_{M,2}}$.
Following GD (1994) we have put $\beta=\beta_0$, where $\beta_0$ is a constant.

(B.5) is the same as equation (4) of GD (1994), so their work is consistent
with this paper for $t>t_{sat}$.  However, GD
(1994) went on to conclude that $\alpha $ saturates 
when $B_1\sim R^{-1/2}_{M,2}
v_2$, apparently assuming from (B.5) that
the criterion for saturation is $R\sim 1$.

However, one must be careful about the second term in (B.5).  Recall that it
is proportional to $H^C_1$, which is constrained
by (B.4).  When one substitutes $H^C_1$ from (B.4) into the second term in
(B.1), one finds that
$$
\alpha = {\alpha_0\over 1+{R^2\over 1+R_{M,2}}} = {\alpha_0 \over 1+{R_{M,2}\over
1+R_{M,2}} \lrp{B_1\over v_2}^2}\; , \dno{7}
$$
so that $\alpha$ saturates not at $B_1\sim R^{-1/2}_Mv_2$, but at
$$
B_{1,\widetilde{\rm sat}} \lrp{1+R_{M,2}\over R_{M,2}}^{1/2} v_2\; , \dno{8}
$$
or, as $R_{M,2}$ gets large
$$
B_{1,\widetilde{\rm sat}} v_2\; , \dno{9}
$$
rather than being resistively limited as  GD suggest.

However, if instead of $\beta=\beta_0$ 
we employ $\beta=\alpha\beta_0/\alpha_0$ (or $q_2=1$ in \pref{40b} and
\pref{41b}), 
then it can be shown analytically 
that the resistively limited 
asymptotic forms $\pref{48}$ and $\pref{49}$ are correct.





\bye